\documentclass[onecolumn,superscriptaddress,amsfonts,amsmath,amssymb,aps,pra,10pt]{revtex4-2}
\usepackage{graphicx}
\usepackage[utf8]{inputenc}
\usepackage[english]{babel}
\usepackage[T1]{fontenc}
\usepackage{amsmath}
\usepackage{amsfonts}
\usepackage{amsmath}
\usepackage{amssymb}
\usepackage{bm}
\usepackage{dsfont} 
\usepackage{changepage}

\newcommand{\bra}[1]{\left<#1\right|} 
\newcommand{\ket}[1]{\left|#1\right>} 
\DeclareMathOperator{\tr}{tr} 
\usepackage{array}
\usepackage{nicematrix}
\usepackage{tikz}
\usetikzlibrary{decorations.pathreplacing}
\usetikzlibrary{calc}

\begin{document}

\title{Designing generalized elegant Bell inequalities \\ in higher dimensions from a Tsirelson bound}

\author{Kwangil Bae}
\email{kibae@kisti.re.kr}
\affiliation{Quantum Network Research Center, Korea Institute of Science and Technology Information (KISTI), Daejeon 34141, Republic of Korea}
\author{Junghee Ryu}
\affiliation{Center for Quantum Information R\&D, Korea Institute of Science and Technology Information (KISTI), Daejeon 34141, Republic of Korea}
\affiliation{Quantum Information, University of Science and Technology, Daejeon, 34113, Republic of Korea}
\author{Ilkwon Sohn}
\author{Wonhyuk Lee}
\affiliation{Quantum Network Research Center, Korea Institute of Science and Technology Information (KISTI), Daejeon 34141, Republic of Korea}

\date{\today}

\begin{abstract}
  Elegant Bell inequality is well known for its distinctive property, being maximally violated by maximal entanglement, mutually unbiased bases, and symmetric informationally complete positive operator-valued measure elements. Despite its significance in quantum information theory demonstrated based on its unique violation feature, it remains the only known one with the characteristic. We present a method to construct Bell inequalities with violation feature analogous to elegant Bell inequality in higher local dimensions from a simple, analytic quantum bound. A Bell inequality with the generalized violation feature is derived in three dimension for the first time. It exhibits larger violation than existing Bell inequalities of similar classes, including the original elegant Bell inequality, while requiring arguably small number of measurements.
\end{abstract}

\maketitle

\section{Introduction}
Violation of a Bell inequality \cite{Bell64,Bell76,Brunner14} implies a fundamental departure from the classical description of correlations between physical systems within local hidden-variable (LHV) theories. The implications of the violation extend well beyond merely confirming conceptual limitations. The correlations violating Bell inequalities enable the certification of physical properties of a system without any assumptions on the internal functioning of the employed devices, a paradigm known as the device-independent (DI) approach \cite{Mayers04,Acin06,Acin07}. The DI approach underlies a range of applications including, DI quantum key distribution \cite{Acin06,Masanes11,Vazirani14,Miller16,Schwonnek21} and DI randomness certification \cite{Pironio10,Miller17,Liu18}. 

Since Tsirelson's seminal contribution setting the strict upper limit for quantum value of Clauser-Horne-Shimony-Holt (CHSH) Bell inequality \cite{Clauser69}, a variety of works have followed to derive analogous bounds for Bell inequalities in generalized scanrios. Although numerical methods based on semidefinite programming provides a versatile way to compute quantum upper bounds within the feasible computational resources \cite{Navascues08}, deriving exact Tsirelson bounds remains challenging even for bipartite case. The exact bounds are known only for comparatively few Bell inequalities among all known ones. To give a few examples, such derivation have been considered in multiple input \cite{Gisin07,Braunstein90,Epping13} and output \cite{Salavrakos17,Kaniewski19,Tavakoli21} scenarios. It is worth noting that the derivation of maximal quantum violation for a broader set of correlation functions rather than for a specific Bell expression can serve as a useful tool for exploring new correlation type Bell inequalities \cite{Epping13,Michalski25}. 

Among the quantum values attainable from a Bell inequality, the maximal one is of particular importance for DI protocols. On the one hand, reaching the largest possible violation typically translates into stronger performance of the protocols, for example, improved tolerance to experimental noise and loss, tighter security parameters \cite{Acin07,Pironio10}. On the other hand, maximal violations corresponding to extremal correlations which, by definition, cannot be expressed as nontrivial convex combinations of other quantum correlations have been favored for DI self-testing \cite{deVincente15}.

Reflecting on the aforementioned possibilities provided by maximal violations of Bell inequalities, it is of prime importance to derive them from desirable features, for example maximal entanglement \cite{Clauser69,Burhman05,Son06,Ji08,Liang09,deVincente15,Salavrakos17,Kaniewski19,Tavakoli21}, mutually unbiased bases (MUBs) \cite{Kaniewski19,Tavakoli21}, or symmetric informationally complete quantum measurements (SICs) \cite{Gisin07,Tavakoli21}. 

As is well known, CHSH Bell inequality is the case for which the maximal violation can be obtained from maximal entanglement and Pauli operators having MUBs as their eigenstates. There exists another bipartite Bell inequality so-called Gisin's elegant Bell inequality (EBI) \cite{Gisin07}, 
\begin{align*}
        &E_{11}+E_{12}-E_{13}-E_{14} \\
        &+ E_{21}-E_{22}+E_{23}-E_{24} \\
        &+ E_{31}-E_{32}-E_{33}+E_{34} \leq  6,
\end{align*}
also optimally violated by maximally entangled state. The expectation value of the product outcomes of Alice's $x$-th measurement and Bob's $y$-th measurement is denoted by $E_{xy}$. Quantum bound for EBI is proved to be $4\sqrt{3}\simeq6.9282>6$ \cite{Acin16}. What makes EBI more interesting is that its violation is obtained from the case where the local measurements of each subsystem is constructed by MUBs and SICs respectively. As first remarked by Gisin, the geometrical symmetries in the optimal measurements of EBI are clearly revealed when they are mapped on Bloch's sphere \cite{Gisin07}. Three eigenstates defining three projective measurements of Alice form three mutually orthogonal vectors on Bloch sphere and Bob's four eigenstates make a tetrahedron \cite{Gisin07}. EBI has been applied for several tasks in quantum information field, including DI randomness certification \cite{Acin16,Andersson18}, DI certification of non-projective measurement \cite{Smania20,Roy23}, and other tasks \cite{Chen21,Tavakoli17,Ghorai18}. Self-testing property of EBI is studied in \cite{Andersson17}. 

MUBs and SICs are not only relevant for the formulation of quantum mechanics \cite{Bengtsson17}, but also for the extensive quantum information processing tasks \cite{Acin16,Andersson18,Shang18,Bae19,Bennett84,Durt10,Ikuta22,Yu08,Tavakoli15,Aguilar18,Wootters89}. Efforts have been made to derive Bell inequalities for MUBs or SICs with higher dimensional system \cite{Kaniewski19,Tavakoli21}. For DI randomness certification, for which higher dimensions provides natural advantage \cite{Acin16}, Bell inequalities for SICs have been used \cite{Tavakoli21,Acin16,Andersson18}. EBI is distinguished from those for either MUBs or SICs as it is optimally violated by both. Randomness certification is a notable application of EBI. DI certification scheme of optimal randomness from one entanglement bit is suggested based on EBI \cite{Acin16,Andersson18}. Considering the above, investigating the desirable feature of Bell inequalities analogous to EBI in higher dimensions is expected to be intriguing from the application perspective beyond its implications just for extending the mathematical symmetries.

We address the problem of extending the violation feature of EBI to higher local dimensional scenario. This investigation is closely related to understanding how the simultaneous consideration of both symmetries in operators, MUBs and SICs, could be advantageous for the violation. Although EBI originated in a different context \cite{Gisin07,Bechmann03}, our focus lies in generalizing its distinctive property of achieving maximal violation through maximal entanglement, MUBs, and SICs. Such generalization of EBI remains largely unexplored so far. Although it does not involve higher dimensions, a generalization of EBI for multiple binary-outome measurements has been recently considered in \cite{Michalski25}. In this work, we derive a simple, tight Tsirelson bound for a functional that generates a family of correlation functions constructed such that they always achieve the maximal value with quantum strategies analogous to the EBI. The main reason for introducing a functional form is to allow for the free choice of SICs, not yet fully characterized in higher dimensions. We expect this work acts as a tool for exploring correlation type Bell inequalities exhibiting the elegant violation feature described above.

This article is structured as follows. A brief introduction on MUBs and SICs, key concepts of our result, are given in "Preliminaries". Bell scenario and relevant notations are defined in "Bell scenario". The generalized violation property inspired from EBI is defined in "Generalized EBI correlation". In "Tsirelson bound", we present the main result deriving the form of correlation function maximized with the generalized EBI correlation. The corresponding quantum bound is derived in an analytic form. To demonstrate the relevance of our result, we derive Gisin's EBI and a novel Bell inequality for three-dimensional generalized EBI correlation in "Derivation of Bell inequality". Therein, we evaluate the violation of the Bell inequality in comparison with established results.

\section{Results}\label{res}
\subsection{Preliminaries} \label{sec_pri}

We briefly introduce the definition of MUBs and SICs, along with their properties relevant to this work. Comprehensive reviews of MUBs and SICs, including their fundamental concepts and properties, can be found in \cite{Durt10,Fuchs17}.\\

\noindent {\bf MUBs.} Normalized bases $\{\ket{a_i}\}_{i=0}^{d-1}$ and $\{\ket{b_j}\}_{j=0}^{d-1}$ are said to be \textit{mutually unbiased} when they satify
\begin{align}
    \left| \left<a_i|b_j\right> \right|^2=\frac{1}{d}
\end{align}
for all combinations of $i$ and $j$. Problem of finding MUBs in all Hilbert space dimesion is still open. However, it is known that whenever the dimension $d$ is prime-power factorizable there are at most $d+1$ MUBs \cite{Wootters89}, referred to as the \textit{complete set}. 

For the case of prime $d$, the complete MUBs can be constructed using Weyl-Heisenberg (WH) group elements. The shift $X:=\sum_{\alpha=0}^{d-1}\ket{\alpha+1}\bra{\alpha}$ and the phase $Z:=\sum_{\alpha=0}^{d-1}\omega^\alpha\ket{\alpha}\bra{\alpha}$ operators generate the WH group, $\{\omega^n X^p Z^q | n,p,q \in [0,d)\}$ with $\omega:=e^{2\pi i/d}$.  The integer domain $\{0,1,\ldots,d-1\}$ is denoted by $[0,d)$ and similar abbreviation is to be used throughout this work. Note that $X^d=Z^d=\mathds{1}$ and the canonical commutation relation (CCR) holds, i.e. $Z^qX^p = \omega^{pq}X^pZ^q$. A subset of WH group without phase factor also forms a group,
\begin{align}
    \mathcal{W}(d)=\{X^m Z^n | m,n \in [0,d)\}.
\end{align}
The complete MUBs can be obtained as the eigenvectors of $d+1$ elements in $\mathcal{W}(d)$, namely $Z$, $X$, $XZ$, \ldots, $XZ^{d-1}$ \cite{Bandyopadhyay02}.

\noindent {\bf SICs.} Suppose a set of $d^2$ normalized states $\{\ket{e_i}\}_{i=0}^{d^2-1}$ satisfying the relation
\begin{align} \label{eq;sic_overlap}
    \left| \left<e_i|e_j\right> \right|^2=\frac{1}{d+1}
\end{align}
for all $i\neq j$. The above constant overlap relations reflects the geometrical symmetry among $d^2$ states. They are equiangular lines in $\mathds{C}^d$ \cite{Fuchs17}. The states define a symmetric informationally complete positive operator-valued measure (POVM), $\left\{\frac{1}{d}\ket{e_i}\bra{e_i}| i \in [0,d^2)\right\}$ with additional condition that all the POVM elements must be summed up to become identity operator. The POVM is \textit{informationally complete} in the sense that its statistics characterizes the quantum state $\rho$ in complex $d$ dimension upon which it is carried on. The $d^2$ number of elements are necessary to define a informationally complete POVM. In this article, a SICs specifically denotes the $d^2$ unit vectors $\{\ket{e_i}\}_{i=0}^{d^2-1}$ satisfying the relation (\ref{eq;sic_overlap}).

It is interesting that SICs are also closely related to WH group. WH operator bases hence provide a suitable common ground for the analysis of MUBs and SICs in higher dimensions. It is the very reason why we adopt the WH bases. SIC-POVM is group covariant if there is a group $\{U_i\}$ having unitary operators as its elements and SICs is generated from operating the elements upon a single vector (or state), so-called fiducial vector $\ket{f}$, i.e. when $\{U_i\ket{f}\}$ is SICs \cite{Renes04}. The group covariance of SICs plays a significant role in the problem of SICs generation \cite{Fuchs17}.

\subsection{Bell Scenario} \label{sec_bel}
\begin{figure}[b]
    \centering
    \includegraphics[scale=0.13]{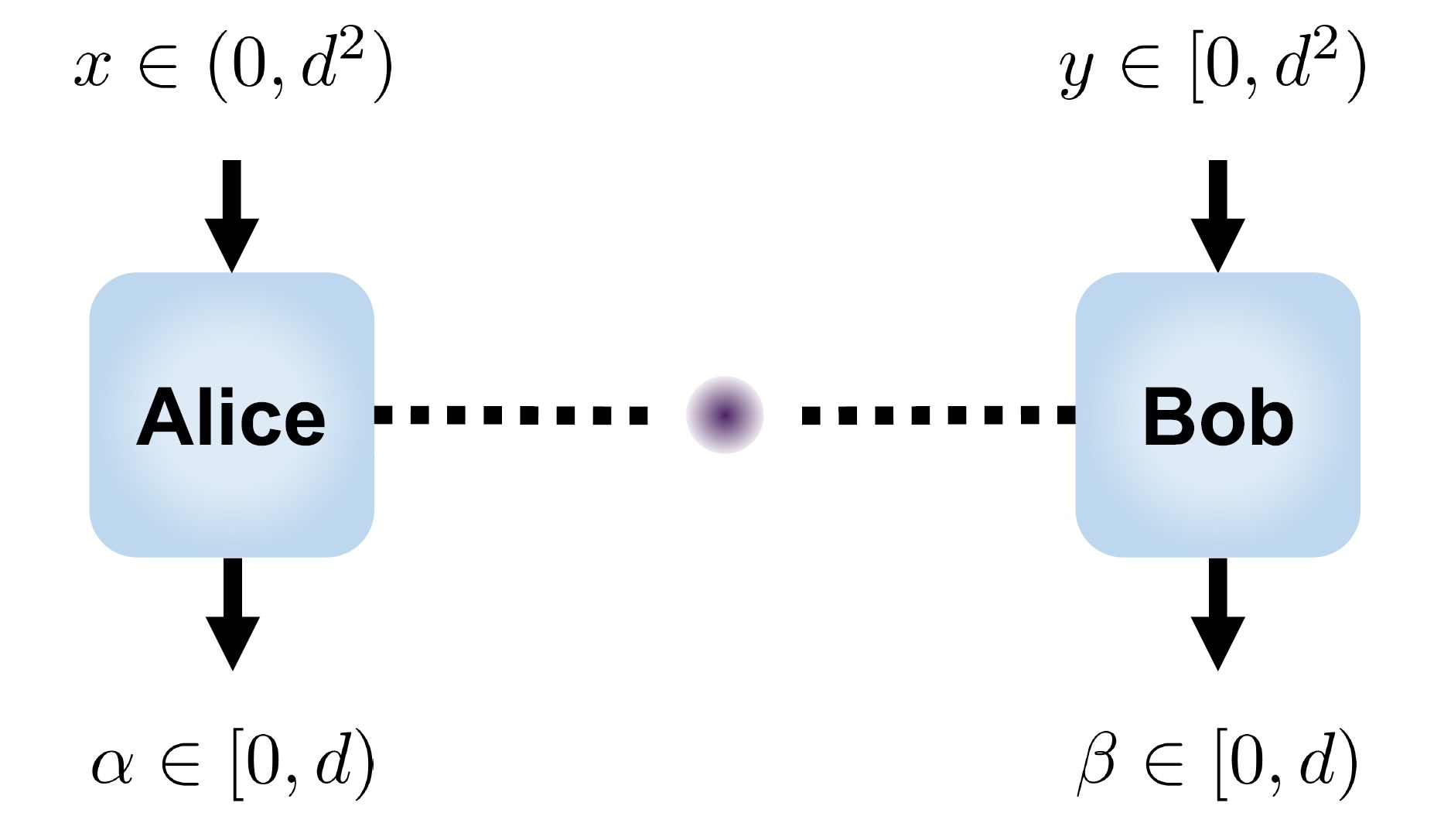}
    \caption{Bell scenario with $d^2-1$ measurement settings for Alice and $d^2$  settings for Bob, each yielding $d$ possible outcomes}
    \label{fig;scenario}
\end{figure}

We consider the measurement scenario of two subsystems Alice (Bob) performing one of possible $d^2-1$ ($d^2$) number of $d$ outcome measurements $A_x$ ($B_y$) on their shared system. Measurements are labelled with $x\in\{1,2,\ldots,d^2-1\}$ and $y\in\{0,1,\ldots,d^2-1\}$. Outcomes of $A_x$ and $B_y$ are respectively distinguished by indices $\alpha \in \{0,1,\ldots,d-1\}$ and $\beta$ defined in the same domain. The scenario is illustrated in Fig. \ref{fig;scenario}. Probability of obtaining the outcomes $\alpha$ and $\beta$ respectively from the choice of $x$ and $y$ is denoted by $p(\alpha\beta | xy)$. A statistics of the Bell test is described by a correlation, $\mathbf{p}:=\{p(\alpha\beta | xy)\}_{\alpha,\beta,x,y}$. Quantum correlation allows
\begin{equation*}
    p(\alpha\beta |xy) = \tr\left[\rho_{AB}(G_\alpha^x\otimes H_\beta^y)\right]
\end{equation*}
where $\rho_{AB}$ is a bipartite density matrix of Alice and Bob. The projectors, $\{G_\alpha^x\}$ and $\{H_\beta^y\}$, are the measurement elements of Alice and Bob respectively. Bell inequality is used to show that there is quantum correlation, which does not allow the local hidden variable (LHV) theoretical description of the correlation \cite{Bell64}. Geometrically, LHV correlation region forms a polytope, $\mathcal{L}$, and quantum region, $\mathcal{Q}$, includes it \cite{Brunner14}.

The generic form of so-called Bell expression \cite{Brunner14} without marginal terms can be defined as
\begin{align}\label{eq;be1}
    S &= \sum_{\alpha,\beta}\sum_{x,y} g_{x,y}^{\alpha,\beta} P(\alpha\beta |xy) \\
      &= \sum_n \sum_{x,y} f_{x,y}^n \langle A_x^n B_y^n \rangle. \label{eq;be2}
\end{align}
where $n\in [1,d)$ and $g_{x,y}^{\alpha,\beta} \in \mathbb{R}$ is the weight for each probability. For simplicity, the summations over all possible values are to be denoted without limits. The correlator $\langle A_x^n B_y^n \rangle = \sum_{\alpha,\beta=0}^{d-1}\omega^{n(\alpha+\beta)}P(\alpha \beta |xy)$, where the outcomes of the measurements $A_x$ and $B_y$ are set as $\omega^\alpha,\omega^\beta \in \{1,\omega,\omega^2,\ldots, \omega^{d-1}\}$. Equivalence of the expressions (\ref{eq;be1}) and (\ref{eq;be2}) are given by the relation $g_{x,y}^{\alpha,\beta} := \sum_n f_{x,y}^n\omega^{n(\alpha + \beta)}$ . The condition $f_{x,y}^{d-n}:=(f_{x,y}^n)^*$ guarantees the real-valuedness of the expression. A possible expression is specified by a $(d^2-1) \times d^2$ coefficient matrices, $F_n$, whose entry $(F_n)_{x,y}$ is defined as $f_{x,y}^n$. We define  $x$-th row and $y$-th column of $F_n$ respectively with column vectors $\mathbf{r}_x^n:=(f_{x,0}^n,f_{x,1}^n,\ldots,f_{x,d^2-1}^n)^\intercal$ and $\mathbf{c}_y^n:=(f_{1,y}^n,f_{2,y}^n,\ldots,f_{d^2-1,y}^n)^\intercal$ such that $F_n=[\mathbf{r}_1^n,\mathbf{r}_2^n,\ldots,\mathbf{r}_{d^2-1}^n]^\intercal =[\mathbf{c}_0^n,\mathbf{c}_1^n,\ldots,\mathbf{c}_{d^2-1}^n]$. The real-valuedness condition can be re-written as $F_n=F_{d-n}^*$, in terms of coefficient matrices. 

 Maximal value of $S$ under LHV theory is $L:=\max_{\mathbf{p}\in\mathcal{L}}S(\mathbf{p})$. Violation of the Bell inequality by quantum mechanics, $L<Q$, is shown with the maximal quantum value,  $Q:=\max_{\mathbf{p}\in\mathcal{Q}}S(\mathbf{p})$. Quantum mechanical expectation can be evaluated with the operator,
\begin{equation}\label{eq;op1}
    \mathcal{B} = \sum_n \sum_{x,y} f_{x,y}^n A_x^n \otimes B_y^n.
\end{equation}
Here, we consider the quantum measurements $A_x=\sum_\alpha \omega^{\alpha}G_\alpha^x$ and $B_y=\sum_\beta \omega^{\beta}H_\beta^y$ where $G_\alpha^x$, $H_\beta^y$ are rank-one projectors.

\subsection{Generalized EBI correlation}\label{sec_gen}

In this section, we define a generalization of the optimal quantum strategies of EBI. The rationale for the definition is also discussed below.

\vspace{1em}
\hrule
\vspace{.5em}
{\bf Definition.} For prime $d\geq 3$, a \textit{generalized EBI correlation} is defined as the bipartite correlation obtained from the quantum realization: 
\begin{adjustwidth}{.5em}{0pt}
\noindent (i) $\ket{\phi_d^+}:=(1/\sqrt{d})\sum_{\alpha=0}^{d-1}\ket{\alpha\alpha}$\\
\noindent (ii) Alice's $d^2-1$ observables comprising  $\mathcal{W}_d \setminus \mathds{1}$ \\
\noindent (iii) Bob's $d^2$ observables respectively having each one of the $d^2$ unit vectors satisfying (\ref{eq;sic_overlap}) as an eigenstate.
\end{adjustwidth}
\vspace{1em}
\hrule
\vspace{1em}

Conditions (i)-(iii) define a possible generalization of optimal quantum realization of EBI. Condition (i) is a two-qudit maximally entangled state. Condition (iii) guarantees the realization of at least one SICs in the Bob's eigenspace. These conditions are straightforwardly deduced from the case of EBI. As for the generalization of Alice's optimal observables, two interpretations are possible. The set of Alice's optimal observables for EBI, $\sigma_x, \sigma_y, \sigma_z$, \cite{Acin16} can be viewed either as ``$d+1$ observables'' defining $d+1$ sets of MUBs or as ``$d^2-1$ unitary operator bases'' \cite{Bengtsson17}. Condition (ii) clearly generalizes the latter. Note that a traceless operator can be expressed as a familiar form $\vec{a}\cdot\vec{\sigma}$ with the unitary operator bases,  $\vec{\sigma}:=(\sigma_x,\sigma_y,\sigma_z)$,  when $d=2$ and  $\vec{a}$ is a three dimensional unit vector. The above duality arises as $d+1=d^2-1$ when $d=2$. Excluding trivial case $X^0Z^0=\mathds{1}$ makes the number of observables $d^2-1$ rather than $d^2$ in condition (ii). Complete MUBs are realized in eigenspace of $d+1$ among $d^2-1$ observables of Alice, namely $Z$, $X$, $XZ$, \ldots, $XZ^{d-1}$ \cite{Bandyopadhyay02}, under condition (ii). Moreover, the eigenstates of all $d^2-1$ observables of Alice falls into one of eigenstates of $Z$, $X$, $XZ$, \ldots, $XZ^{d-1}$ under condition (ii). 
In other words, the eigenspaces of $d^2-1$ observables are $d-1$ folded eigenspaces of $d+1$ complete MUBs. (See Appendix \ref{app_mub} for the detailed discussion.)

WH operators play a pivotal role in not only deriving Alice's optimal observable but also for Bob's ones. The explicit derivation is given in the following section. Here, we briefly convey the advantages brought by exploiting WH group covariant SICs for satisfying condition (iii). For brevity, consider the parameterization $W_{dp+q}:=X^pZ^q$ such that $\mathcal{W}(d) = \{W_j | j \in [0,d^2) \}$. For example, $W_5=XZ^2$ when $d=3$. With the parameterization, WH group covariant SICs are $\{W_j \ket{\varphi} | j \in [0,d^2) \}$ for a fiducial state $\ket{\varphi}$. Group covariant SICs enable one to satisfy the condition (iii) in a \textit{systematic} way. The condition is satisfied by assigning each one of $d^2$ states, $W_j \ket{\varphi}$, to a different eigenstate of Bob's $j$-th measurement. Consequently, Bob's optimal measurements defined in condition (iii) are derived neatly as (\ref{eq;bob1}), in the following section. The WH group covariant SICs also provide \textit{versatile nature} for our framework. It is based on the fact that our method for deriving the quantum bound can always be applied if WH group covariant SICs of given dimension exists. The existence of SICs is known in all dimension up to $d=193$ \cite{DeBrota21}. More importantly, all group covariant SICs are covariant to WH group in prime dimension \cite{Zhu10} that we mainly consider. In light of the above discussion, exploiting WH group covariance of SICs is arguably reasonable to derive and analyze the Bell nonlocality for \textit{generalized EBI correlation}. 

In the following section, we elaborate on how the choice of WH operator bases also offer advantages in deriving optimal quantum value. 

\subsection{Tsirelson bound}\label{app_tsirelson}
In this section, we suggest the correlation function for which tight upper bound of their quantum values is obtained from generalized EBI correlation, if only the WH group covariant SICs is defined in the considered dimension. Sum-of-squares (SOS) decomposition technique is used for the derivation (see Appendix \ref{app_qua}).

Consider the shifted operator, $\bar{\mathcal{B}} := Q \mathds{I} - \mathcal{B}$ and operators $P_y^n:= D_y^n \otimes \mathds{1} - \mathds{1} \otimes (B_y^n)^\dagger$ and $D_y^n:=\sum_x f_{x,y}^n A_x^n$, where $Q \in \mathbb{R}$ is a real constant and $\mathds{1}$ is a $d \times d$ identity matrix. The identity matrix acting on the composite system is denoted by $\mathds{I}:=\mathds{1}\otimes \mathds{1}$. The SOS decomposition $\bar{\mathcal{B}}=\sum_{n,y} (P_y^n)^\dagger P_y^n \geq  0$ directly implies $\left<\mathcal{B}\right>\leq Q$.  Substituting $P_y^n$ into the decomposition derives
\begin{align}\label{eq;sos}
     \frac{1}{2}  \sum_{n,y}(D_y^n)^\dagger D_y^n \otimes \mathds{1} + \frac{1}{2}  \sum_{n,y} \mathds{1} \otimes B_y^n (B_y^n)^\dagger = Q\mathds{I}.
\end{align}

\noindent In the derivation, the Hermiticity of $\mathcal{B}$, i.e. $\mathcal{B} = \sum_{n,y} D_y^n\otimes B_y^n=\sum_{n,y} (D_y^n)^\dagger \otimes (B_y^n)^\dagger$, is used. Hereafter, the Eq.(\ref{eq;sos}) is to be referred to as `decomposition condition'. We define `saturation condition' as $\tr[\bar{\mathcal{B}}\rho]=0$ for any valid quantum strategies. It implies that $Q$ is attainable from quantum mechanics. It is known that $P_y^n\rho=0 \,\,\forall \,y,n$  is equivalent to the saturation condition \cite{Kaniewski19}. Importantly, this condition is satisfied by $D_y^n=(B_y^n)^*$ on a maximally entangled state as $P_y^n\ket{\phi_d^+}=[D_y^n\otimes \mathds{1} - \mathds{1} \otimes (B_y^n)^\dagger]\ket{\phi_d^+}=[D_y^n - (B_y^n)^*] \otimes \mathds{1}\ket{\phi_d^+}$. An identity, $M\otimes \mathds{1}\ket{\phi_d^+} = \mathds{1}\otimes M^\intercal \ket{\phi_d^+} \forall M$, is used in the second equation. This condition reveals the relation between optimal measurements on both subsystems.

Our strategy is to solve the decomposition and saturation conditions simultaneously by varying the coefficients of Bell expression $f_{x,y}^n$ with a given generalized EBI correlation. To this end, we define the optimal measurements obeying conditions (ii) and (iii),
\begin{align}
    A_{dp+q} &= \tau W_{dp+q} \quad \, \forall p,q \label{eq;alice} \\ 
    B_y &= W_y B_0 W_y^\dagger \quad \forall y \label{eq;bob1}
\end{align}
where $\tau:=\omega^{pq \delta(d,2)/2}$ \cite{Note} is a constant defined with Kronecker delta, $\delta$. The parameterization $x(p,q):=dp+q$ is used in Eq.(\ref{eq;alice}). It allows one to express Alice's measurements with a single measurement index, $x$. Measurements are specified with $x \in (0,d^2)$ or $p,q \in[0,d)$ satisfying $(p,q)\neq (0,0)$ and $y \in [0,d^2)$. Alice's measurements in (\ref{eq;alice}) are given as $d^2-1$ WH operators when $d \geq 3$. Bob's measurements are generated from $B_0$ as  Eq.(\ref{eq;bob1}). An important part of our method is to construct $B_0$ such that it has at least one fiducial state as its eigenstate. This choice of $B_0$ makes it possible to realize WH group covariant SICs on Bob's eigenspace. Consider $B_0=\sum_\beta \omega^\beta \ket{\varphi_\beta}\bra{\varphi_\beta}$ whose eigenspace is defined as $\Phi_d:=\{ \ket{\varphi_\beta} | \beta\in[0,d) \}$, a set of $d$ orthonormal states including the fiducial state $\ket{\varphi}$. Without loss of generality, one can let $\ket{\varphi}:=\ket{\varphi_{d-1}}$.  Then we have,
\begin{align}\label{eq;bob2}
    B_y&=\sum_\beta \omega^\beta W_y\ket{\varphi_\beta}\bra{\varphi_\beta}W_y^\dagger
\end{align}
where $\ket{\varphi}:=\ket{\varphi_{d-1}}$. The above spectral decomposition shows that the eigenbases of $B_y$ are now $\{ W_y\ket{\varphi_\beta} | \beta\in[0,d) \}$. Consequently, as $W_y\ket{\varphi_{d-1}}= W_y\ket{\varphi}$, collecting the set of $(d-1)$-th eigenstates of all $d^2$ Bob's measurements, WH group covariant SICs $\{ W_y\ket{\varphi} | y\in[0,d^2) \}$ is obtained. Note that conditions (ii) and (iii) are satisfied with the choice (\ref{eq;alice}) and (\ref{eq;bob2}).

First, we consider the decomposition condition. The second term in the left-hand side (LHS) of (\ref{eq;sos}) reduces to $d^2(d-1)\mathds{I}/2$ as $B_y^n$ is unitary. The first term can be converted to $\sum_{n,y}(D_y^n)^\dagger D_y^n=\sum_n \lVert F_n\rVert^2 \mathds{I} +\sum_n \sum_{a\neq x} (\mathbf{r}_a^n)^* \cdot \mathbf{r}_x^n(A_a^n)^\dagger A_x^n$ \cite{SM} where $\lVert \cdot \rVert$ is the Frobenius norm and $\mathbf{a}\cdot \mathbf{b} := \mathbf{a}^\intercal \mathbf{b}$. The LHS then results in a desirable form $\tilde{Q}\mathds{I}$ with
\begin{equation}\label{eq;qbound}
\tilde{Q}(d) :=\frac{1}{2} \left[ \sum_n \lVert F_n\rVert^2 + d^2(d-1) \right] 
\end{equation}
when $(\mathbf{r}_a^n)^* \cdot \mathbf{r}_x^n=0$ for $a \neq x$. 

Second, we simplify the saturation condition with generalized EBI correlation, namely with correlation from $\ket{\phi_d^+}$ and measurements (\ref{eq;alice}) and (\ref{eq;bob2}). As previously mentioned, the saturation condition can be satisfied with the maximally entangled state and  $(B_y^n)^* = D_y^n = \sum_x f_{x,y}^n A_x^n = \mathbf{c}_y^n \cdot \mathbf{A}_n$  $\forall y,n$ where $\mathbf{A}_n:=(A_1^n,A_2^n,\ldots,A_{d^2-1}^n)^\intercal$. In the derived equation, the only variable parameters are $\mathbf{c}_y^n$, or equivalently $f_{x,y}^n$. Considering two conditions altogether, the above SOS decomposition implies that (\ref{eq;qbound}) is the Tsirelson type bound of correlation function $S$ defined with $\{f_{x,y}^n\}$ satisfying $(\mathbf{r}_a^n)^* \cdot \mathbf{r}_x^n=0$ and $(B_y^n)^* =\mathbf{c}_y^n \cdot \mathbf{A}_n$ for all $x \neq a, y,n$. The solution is derived in the following theorem. 

{\bf  Theorem.} Consider prime dimension $d$, in which WH group covariant SICs is defined. Then, there always exists the expression $S$ defined with
\begin{equation}\label{eq;gee}
    f_{dp+q,dr+s}^n = \omega^{-n(ps+qr)} f_{dp+q,0}^n
\end{equation}
for $p,q,r,s \in [0,d)$, $dp+q\neq0$ satisfying $Q=\tilde{Q}$ with \textit{generalized EBI correlation}.

\textit{Proof.}---Consider an optimal measurement $B_0=\sum_{\beta=0}^{d-1} \omega^\beta \ket{\varphi_\beta}\bra{\varphi_\beta}$ whose eigenspace contains a fiducial state $\ket{\varphi}_{d-1}:=\ket{\varphi}$. This choice of $B_0$ realizes SICs in the eigenspace, as previously explained. Applying saturation condition for $y=0$, $(B_0^n)^* = \mathbf{c}_0^n \cdot \mathbf{A}_n$, into the RHS of Eq.(\ref{eq;bob1}), it is derived that
\begin{align}
(W_y B_0^n W_y^\dagger)^* &= \mathbf{c}_0^n \cdot (W_y^* \mathbf{A}_n W_y^\intercal) \nonumber \\
                           &= \sum_{p,q} \omega^{-n(ps+qr)} f_{dp+q,0}^n A_{dp+q}^n. \label{eq;proof_RHS}
\end{align}
Substituting $A_{dp+q}^n=(\tau X^p Z^q)^n$ and $W_y=W_{dr+s}=X^rZ^s$ into the RHS of the first equation and suitably applying the canonical commutation relation, $Z^qX^p = \omega^{pq}X^pZ^q$, derives Eq.(\ref{eq;proof_RHS}). The parameterization $x=dp+q$ and $y=dr+s$ are used in the derivation. While, the saturation condition for arbitrary $y$, $(B_y^n)^* = \mathbf{c}_y^n \cdot \mathbf{A}_n$, reads
\begin{align}
    (B_y^n)^* &=\mathbf{c}_y^n \cdot \mathbf{A}_n \nonumber \\
              &=\sum_x f_{x,y}^n A_x^n \nonumber \\
              &=\sum_{p,q} f_{dp+q,dr+s}^n A_{dp+q}^n \label{eq;proof_LHS}
\end{align}
where the same parameterization $x=dp+q$ and $y=dr+s$ are used for comparison with (\ref{eq;proof_RHS}). Consequently, one obtain the claimed Eq.(\ref{eq;gee}) from equating (\ref{eq;proof_RHS}) and (\ref{eq;proof_LHS}). The expression given as (\ref{eq;gee}) also satisfies the decomposition condition. Substituting (\ref{eq;gee}) into $(\mathbf{r}_a^n)^* \cdot \mathbf{r}_x^n = \sum_y (f_{a,y}^n)^*f_{x,y}^n$, one finds that it reduces to 0 for all $n$ if $a \neq x$. Showing saturation condition is always solvable completes the proof.  The elements of $\mathbf{A}_n$ is the unitary operator bases for traceless operator \cite{Bengtsson17}. It means that the traceless operator $(B_0^n)^*$ can always be expressed  as a linear combination form $\mathbf{c}_0^n \cdot \mathbf{A}_n$ with suitable $\mathbf{c}_0^n$.  \null\hfill $\blacksquare$ 

The above theorem shows that the Bell expression maximized by \textit{generalized EBI correlation} can always be defined with (\ref{eq;gee}). It is worth noting that one has freedom in choice of SICs which determine $f_{x,0}^n$, the variable parameters in the RHS of (\ref{eq;gee}). With a chosen WH covariant SICs and the corresponding fiducial state, one can construct the optimal $B_0$ as previously explained. Then the coefficients for $y=0$, $f_{x,0}^n$, are derived from the saturation condition, $(B_0^n)^* =\sum_x f_{x,0}^n (\mathbf{A}_n)_x$. Different Bell expressions can be generated by varying $f_{x,0}^n$ in the RHS of Eq.(\ref{eq;gee}) with different choices of SICs. We formally summarize the above statement. The functional that is maximized by the generalized EBI correlation is written as

\begin{align}\label{eq;set_corr}
     S_d:=\sum_{n=1}^{d-1} \sum_{\substack{p,q,r,s=0 \\ dp+q\neq 0}}^{d-1} \omega^{-n(ps+qr)} f_{dp+q,0}^n \langle A_x^n B_y^n \rangle.
\end{align}

\noindent From the functional, a Bell expression can be derived by substituting $f_{dp+q,0}$, which is the solution of $(B_0^n)^* = (\sum_\beta \omega^{n\beta} \ket{\varphi_\beta}\bra{\varphi_\beta})^* = \sum_x f_{x,0}^n W_x^n$, with different cohices of eigenbases $\{\ket{\varphi_\beta}\}_{\beta=0}^{d-1}$ generated by a chosen fiducial vector, $\ket{\varphi}=\ket{\varphi_{d-1}}$.

We also note here that assuming Alice's measurements as unitary operator bases \cite{Bengtsson17} for traceless operator is advantageous for restricting the number of measurements in our method. The traceless unitary operator bases $\{U_x\}$ are the minimum number of unitary operators that satisfy $V =\sum_x v_x  U_x$ for an arbitrary traceless unitary operator $V$. Therefore, with our choice of Alice's measurements, $B_0$ constructed from any fiducial vector can be considered to satisfy the saturation condition, $(B_0^n)^* =\sum_x f_{x,0}^n (\mathbf{A}_n)_x$. In other words, any WH covariant SICs can be considered. As a Bell test requires random inputs to close freedom-of-choice loophole \cite{Bell64}, reducing the number of settings can be advantageous for lowering input randomness. Also, fewer measurements might beneficial to restrict the number of experimental parameters to be controlled. 

\subsection{Derivation of Bell inequality} \label{sec_der}
The above results may provide a useful tool for deriving novel Bell inequalities in higher dimensions. First, Bell expressions can be studied within the restricted set generated from (\ref{eq;set_corr}) whose quantum upper bound is guaranteed to be obatained from generalized EBI correlation. Second, the derived Tsirelson bound can be used to derive the Bell inequality, by providing a quanm value to be compared with LHV model. In this section, we briefly illustrate the above possiblity with examples.

\subsubsection{Gisin's elegant Bell inequality} \label{sec_gis}

Gisin's elegant Bell inequality can be derived from a $d=2$ case of our method. We choose optimal $B_0=\ket{\varphi_0}\bra{\varphi_0}-\ket{\varphi_1}\bra{\varphi_1}$ with known fiducial states $\ket{\varphi_0}=\mathcal{N}_0 (e^{-\pi i/4},\sqrt{2}/(1-\sqrt{3}))^\intercal$ and $\ket{\varphi_1}=\mathcal{N}_1 (e^{-\pi i/4}, \sqrt{2}/(1+\sqrt{3}))^\intercal$ \cite{Renes04} satisfying $\langle \varphi_0 | \varphi_1 \rangle=0$. $\mathcal{N}_0$ and $\mathcal{N}_1$ are the normalization constants. From the saturation condition of the SOS decomposition, $B_0^* = \mathbf{c}_0 \cdot \mathbf{A}$, the solution $\mathbf{c}_0 = (f_{1,0},f_{2,0},f_{3,0})^\intercal=(1/\sqrt{3})(-1,-1,1)^\intercal$ is easily derived with $\mathbf{A}=(Z,X,iXZ)^\intercal=(\sigma_z,\sigma_x,\sigma_y)^\intercal$. As there is only one case of $n=1$, the indices denoting it are omitted. Substituting $\mathbf{c}_0$ into Eq.(\ref{eq;gee}), coefficients matrix is derived as $F_1:=1/\sqrt{3}\left[(-1,-1,1,1),(-1,1,-1,1),(1,-1,-1,1)\right]$. The coefficients correspond to those of original EBI up to a multiplicative constant $1/\sqrt{3}$ and re-labelling of $y$, specifically $y\rightarrow 3-y$. With coefficients $F_1$, the Bell inequality is derived with $L=2 \sqrt{3}$  \cite{Acin16}. Quantum violation $\tilde{Q}=4$ can be evaluated from (\ref{eq;qbound}) with $\lVert F_1 \rVert =2$. 

\subsubsection{Bell inequality for two-qutrit system} \label{sec_ine}

We derive a Bell inequality maximally violated by generalized EBI correlation for locally three dimensional system. For $d=3$, there are infinitely many fiducial states generating WH covariant SICs, for example $\ket{\psi} = 1/\sqrt{2}(0,1,-\omega^{\theta})$ for $0\leq\theta\leq1/2$ \cite{Tabia13}. We choose one with $\theta=0$, $\ket{\varphi}=1/\sqrt{2}(0,1,-1)=1/\sqrt{2}\left(\ket{1}-\ket{2}\right)$ which is known to generate the Hesse SICs \cite{Fuchs17}. With the set of orthonormal bases $\ket{\varphi_0}:=\ket{0}$, $\ket{\varphi_1}:=1/\sqrt{2}(\ket{1}+\ket{2})$, and $\ket{\varphi_2}:=\ket{\varphi}$, optimal $B_0$ is constructed as
\begin{align}
B_0=\sum_\beta \omega^\beta \ket{\varphi_\beta}\bra{\varphi_\beta}=
\begin{pmatrix}
        1 & 0 & 0\\
        0 & -1/2 & i\sqrt{3}/2\\
        0 & i\sqrt{3}/2 & -1/2
\end{pmatrix}.
\end{align}
\noindent It is worth mentioning here that our choice of $\ket{\varphi_1}$ is also a fiducial state for another WH covariant SICs. By solving the saturation condition, $(B_0)^* = \mathbf{c}_0^1\cdot \mathbf{A}_1$, the first column of $F_1$ is straightforwardly obtained as $\mathbf{c}_0^1=(f_{1,0}^1,f_{2,0}^1,\ldots,f_{8,0}^1)^\intercal=(1/2)(1,1,\lambda,\nu,\mu,\lambda,\mu,\nu)^\intercal$ with $\lambda:=-i/\sqrt{3}$, $\mu:=\omega \lambda$ and $\nu:=\omega^2 \lambda$. Substituting $\mathbf{c}_0$ into Eq.(\ref{eq;gee}) and using the real-valuedness condition $F_2=F_1^*$, the coefficients are fully generated as presented in \cite{SM}. The derived coefficients for the first column can be further simplified in a functional form,
\begin{align}\label{eq;S3}
    f_{dp+q,0}^n=
    \begin{cases}
    \frac{1}{2}, \quad (p=0) \\
    \frac{i}{2\sqrt{3}}(-1)^n\omega^{-npq}, \quad (p\neq0).
    \end{cases}
\end{align}

\noindent Substituting the derived coefficients (\ref{eq;S3}) into (\ref{eq;gee}), a Bell expression is defined as:
\begin{align}\label{eq;bi}
S_3 := \sum_{n=1}^2\sum_{\substack{p,q,r,s=0\\ 3p+q \neq0}}^2 \omega^{-n(ps+qr)} f_{3p+q,0}^n\left\langle A_{3p+q}^n B_{3r+s}^n \right\rangle.
\end{align}
A Bell inequality for generalized EBI correlation in two-qutrit system is derived as $S_3\leq 15$. The upper bound $L=15$ can be obtained in the set of local deterministic correlations obeying $P(\alpha \beta|xy)=1$ for a certain combination of $\alpha,\beta$ and otherwise the probability is 0. The optimal strategy is found as, for any $x$ and respectively for $y=0,1,2$, $y=3,6$, and $y=4,5,7,8$, assigning $P(00|xy)$, $P(01|xy)$, $P(02|xy)$ as 1. 

Violation is obtained from the maximal quantum value, $\tilde{Q}=18$, evaluated from (\ref{eq;qbound}) with $\lVert F_1 \rVert= \lVert F_2 \rVert  =9$. It is obtained by two parties performing measurements, $A_x=W_x$ and $B_y=W_yB_0W_y^\dagger$, on a two-qutrit maximally entangled state $|\phi^+_3\rangle$ when $x \in (0,9)$ and $y \in [0,9)$. As we considered two fiducial states, $\ket{\varphi_1}$ and $\ket{\varphi_2}$, $2 d^2$ eigenstates on Bob's side $\{W_y\ket{\varphi_1}|y \in [0,9)\}$ and $\{W_y\ket{\varphi_2}|y \in [0,9)\}$ make dual SICs. The eigenstates of Alice's four optimal observables for $x=1,3,4,5$, namely $Z,X,XZ,XZ^2$, define the complete MUBs \cite{SM}. 

We remark that the derived Bell inequality $S_3$ shows stronger violation than known Bell inequalities of similar classes. The critical visibility $\nu^c$, the infimum of visibilities $\nu \in [0,1]$ in the isotropic state $\rho=\nu |\phi_d^+ \rangle \langle\phi_d^+|+(1-\nu)\mathds{I}/d^2$ for which a Bell inequality is violated, is 83.33\% for $S_3$. It is smaller than that of Bell inequalities defined for $d=3$ MUBs, 96.77\% \cite{Kaniewski19}, 96.63\% \cite{Tavakoli21}, SICs, 96.41\% \cite{Tavakoli21}, and original EBI, 86.6\%. All the above critical visibility values are converted to numerical values and rounded to two decimal places for comparison. In addition, the number of settings required for Alice is 8 for $S_3$ and it is smaller than the case of known Bell inequality for $d=3$ SICs which requires $\binom{d^2}{2}=36$ settings for Alice when the number of Bob's setting is same as 9 \cite{Tavakoli21}.

\section{Discussion}
We present a bipartite correlation function for which the maximal quantum value can always be achieved with maximal entanglement, mutually unbiased bases (MUBs), and symmetric informationally complete measurement elements (SICs). The violation property is inspired from that of Gisin's elegant Bell inequality (EBI). We also derive the tight quantum upper bound in analytic form. The correlation function is tailored to depend on a choice of SICs, enabling a systematic search over candidate SICs whose explicit constructions remain incomplete in higher dimensions. That all WH group covariant SICs can be considered highlights the versatility of our result. We especially note that for the prime dimensions considered here, all known SICs are WH group covariant. It is also discussed that the number of measurement settings are restricted while maintaining this versatility. 

Our primary contribution lies in proposing a method to generate Tsirelson-type inequalities for the violation property described above. The exploration of the corresponding Bell inequalities in higher dimensions represents an interesting direction for future research, requiring broader analysis such as finding suitable choices of SICs in each dimension. We expect that our result will facilitate such exploration. In our framework, deriving a specific Bell inequality reduces to deriving a Bell expression from the suggested functional with a suitable SICs and comparing LHV bound with the corresponding Tsirelson bound, provided by this work. Consequently, one does not have to explore full space of Bell expressions.

To showcase the potential of our result, we derived the original Gisin's EBI and a novel Bell inequality for two-qutrit system within our framework, corresponding to two and three dimensional cases, respectively. Notably, the derived inequality exhibits substantially larger violation than existing Bell inequalities, either for three dimensional MUBs \cite{Kaniewski19,Tavakoli21} or SICs \cite{Tavakoli21}. The critical visibility of isotropic state for the derived inequality is more than 13\% lower for all considered cases. Moreover, the critical visibility is smaller compared to Gisin's EBI, also. These findings establishes a natural foundation for future research on self-testing MUBs and SICs in higher dimensional systems, as well as DI quantum cryptographic protocols which would be built upon them.

\begin{acknowledgments}
This research was supported by Korea Institute of Science and Technology Information (KISTI) (No. K26L1M3C5) and the National Research Council of Science \& Technology (NST) grant by the Korea government (MSIT) (No. CAP22053-000). J.R. acknowledges support from the National Research Foundation (NRF) of Korea grant funded by the Korea Government (Grant No. RS-2023-NR119931).
\end{acknowledgments}

\appendix
\newpage

\section{Complete MUBs in Alice's eigenspace}\label{app_mub}

Complete MUBs are realized in the eigenspace of Alice's $d^2-1$ measurements as stated in condition (ii) of generalized EBI correlation. The existence of complete MUBs is straightforwardly shown from $ \mathcal{C} := \{ Z, X, XZ, \ldots, XZ^{d-1} \} \subset \mathcal{W}_d$ and the fact that the eigenstates of the $d+1$ unitary operators in $\mathcal{C}$ are complete MUBs for the case of prime $d$. We prove another notable property regarding MUBs in Alice's eigenspace in the following proposition. \\

\noindent \textbf{Proposition.} The collection of all eigenstates of Alice's measurements defined in \textit{generalized EBI correlation} are a $(d-1)$-fold complete MUBs.\\

\noindent \textit{Proof.} The measurements generating complete MUBs, $\mathcal{C}$, are divided into two subsets $\mathcal{C}_1:=\{Z\}$ and $\mathcal{C}_2:=\{ XZ^q | q\in[0,d) \}$ such that $\mathcal{C}=\mathcal{C}_1\cup \mathcal{C}_2$. The set of Alice's measurements $\mathcal{A}:=\{ X^pZ^q | (p,q)\neq (0,0) \}$ can be similarly divided with $\mathcal{A}_1:=\{Z^q | q\neq0 \}$ and $\mathcal{A}_2:=\{X^pZ^q | p \neq 0 \}$, i.e. $\mathcal{A}=\mathcal{A}_1\cup \mathcal{A}_2$. 

Then the above proposition is proved by
\begin{align}
    V(\mathcal{C}_1)&=V(\mathcal{A}_1), \,\mbox{and} \label{eq;prop1}\\
    V(\mathcal{C}_2)&=V(\mathcal{A}_2) \label{eq;prop2_1}
\end{align}
where $V(O)$ denotes a set of unit eigenvectors of all elements in a given set, $O$. The Eq.(\ref{eq;prop1}) are easily proved, since taking the power of a unitary operator preserves the eigenspace. 

Applying the same property of unitary matrices to the left-hand side (LHS) of the Eq.(\ref{eq;prop2_1}) derives $V(\mathcal{C}_2)=V(\{ (XZ^q)^p | p \neq 0 \})$. It also holds that $(XZ^q)^p=\omega^n X^pZ^{\dot{(pq)}}$ from the canonical commutation relation, $Z^qX^p = \omega^{pq}X^pZ^q$, where $n(p,q):=qp(p-1)/2$ is an integer and the overdot denotes the smallest non-negative $d$ modulo value. Then, one obtains $V(\mathcal{C}_2)=V(\{ X^pZ^{\dot{(pq)}} | p \neq 0 \})$. Consequently, the Eq.(\ref{eq;prop2_1}) can be rewritten as 
\begin{align} \label{eq;prop2_2}
    V(\{ X^pZ^{\dot{(pq)}} | p \neq 0 \})=V(\mathcal{A}_2)=V(\{X^pZ^q | p \neq 0 \}).
\end{align}
This equation can be proved via the following equivalence: 
\begin{align} \label{eq;equiv1}
    q_1 \neq q_2 \Leftrightarrow \dot{(pq_1)} \neq \dot{(pq_2)} 
\end{align}
for $p \in (0,d)$ and $q_1,q_2 \in [0,d)$. It means that all different values of $q$ ($q_1$ and $q_2$) implies different values of $\dot{(pq_1)}$ and $\dot{(pq_2)}$. The situation results in the one-to-one correspondence between set of $q$ and set of $\dot{(pq)}$ for any given non-zero $p$. This is sufficient to prove Eq.(\ref{eq;prop2_2}).

We prove the rightward relation ($\Rightarrow$) first. The least non-negative residue of an integer $x$ modulo-$d$ is given by $\dot{x}=x-d\lfloor x \rfloor$ where $\lfloor x \rfloor$ denotes the corresponding quotient. Using this expression, the RHS of (\ref{eq;equiv1}) can be recast as
\begin{align}\label{eq;equiv2}
    \frac{p(q_1-q_2)}{d} \neq  \left\lfloor \frac{pq_2}{d} \right\rfloor - \left\lfloor \frac{pq_1}{d} \right\rfloor.
\end{align}
Consider now the LHS of (\ref{eq;equiv1}). Without loss of generality, we assume $q_1>q_2$. Then, as $d$ is prime and $p$ and $q_1-q_2$ are all smaller than $d$, the LHS of (\ref{eq;equiv2}) is always a fractional number. While the RHS clearly is an integer. It proves (\ref{eq;equiv2}), thus rightward relation. The leftward relation ($\Leftarrow$) follows directly from its contrapositive, $q_1 = q_2 \Rightarrow \dot{(pq_1)} = \dot{(pq_2)}$. \\
\null\hfill $\blacksquare$\\

 We note that the above proposition implies that any one of the eigenstates of Alice's measurements fall into one of the eigenstates of $ Z, X, XZ, \ldots, XZ^{d-1}$ that comprise the complete MUBs.

\section{Derivation of Tsirelson bound} \label{app_qua}
In this section, we supplement the derivation of quantum upper bound $\tilde{Q}$ with additional details. As shown in the main text ``Tsirelson bound", the SOS decomposition $\bar{\mathcal{B}}=\sum_{n,y} (P_y^n)^\dagger P_y^n \geq  0$ is modified to
\begin{align}\label{s.eq;sos}
     \mathcal{B} \leq \frac{1}{2}  \sum_{n,y}(D_y^n)^\dagger D_y^n \otimes \mathds{1} + \frac{1}{2}  \sum_{n,y} \mathds{1} \otimes B_y^n (B_y^n)^\dagger 
\end{align}

\noindent by substituting $P_y^n:= D_y^n \otimes \mathds{1} - \mathds{1} \otimes (B_y^n)^\dagger$. The second term of the RHS of  (\ref{s.eq;sos}) reduces to $d^2(d-1)/2 \mathds{I}$ by the unitarity of $B_y^n$. To convert the first term of the RHS into a real scalar multiple of the identity, it suffices to substitute $D_y^n:=\sum_x f_{x,y}^n A_x^n$ into $\sum_{n,y}(D_y^n)^\dagger D_y^n$, showing that,
\begin{align}\label{s.eq;derivation}
    \sum_{n,y}(D_y^n)^\dagger D_y^n &=\sum_{n,y} \left( \sum_a f_{a,y}^n A_a^n \right)^\dagger \sum_x f_{x,y}^n A_x^n     \nonumber \\
    &= \sum_{n,a,x} \left[ \sum_y (f_{a,y}^n)^* f_{x,y}^n \right] (A_a^n)^\dagger A_x^n  \nonumber \\
    &= \sum_{n,x,y} |f_{x,y}^n|^2 \mathds{1}+\sum_n\sum_{a\neq x} \left[ \sum_y (f_{a,y}^n)^* f_{x,y}^n \right] (A_a^n)^\dagger A_x^n \nonumber \\
        &= \sum_n \lVert F_n\rVert^2 \mathds{1} +\sum_n \sum_{a\neq x} (\mathbf{r}_a^n)^* \cdot \mathbf{r}_x^n (A_a^n)^\dagger A_x^n.
\end{align}
where $\lVert \cdot \rVert$ is the Frobenius norm and $\mathbf{a}\cdot \mathbf{b} := \mathbf{a}^\intercal \mathbf{b}$. Consequently, with $(\mathbf{r}_a^n)^* \cdot \mathbf{r}_x^n=0$ for all $n,a\neq x$, the upper bound in inequality (\ref{s.eq;sos}) simplifies to
\begin{equation}\label{s.eq;qbound}
\tilde{Q} :=\frac{1}{2} \left[ \sum_n \lVert F_n\rVert^2 + d^2(d-1) \right] .
\end{equation}

\section{Bell expression and optimal measurements}\label{app_beo}
For the derived Bell inequality $S_3$, we present the explicit form of coefficient matrix, MUBs and SICs that correspond to maximal violation. Although the symmetries of $S_3$ can be expressed analytically as in the main text, their realization in the the coefficient matrix, MUBs and SICs makes these symmetries apparent at a glance. To this aim, considering coefficients in $F_1$ is sufficient as all the other coefficients is obtained from the realvaluedness condition, $F_2=F_1^*$. It is obtained as $8\times 9$ matrix,\\

\begin{equation}
F_1 = \frac{1}{2}
\begin{pNiceMatrix}[margin]
    \Block[fill=blue!10]{8-1}{}   1 & 1 & 1 & \Block[fill=black!10]{2-3}{} \omega^2 & \omega^2 & \omega^2 & \omega & \omega & \omega  \\ 
        1 & 1 & 1 & \omega & \omega & \omega & \omega^2 & \omega^2 & \omega^2  \\ \Block[fill=blue!25]{3-1}{}
        \lambda & \Block[fill=black!10]{3-2}{} \nu & \mu & \lambda & \nu & \mu & \Block[fill=black!10]{3-3}{} \lambda & \nu & \mu \\
        \nu & \mu & \lambda &  \mu & \lambda & \nu &  \lambda & \nu & \mu \\
        \mu & \lambda & \nu &  \nu & \mu & \lambda &  \lambda & \nu & \mu \\ 
        \lambda & \mu & \nu & \Block[fill=black!10]{3-3}{} \lambda & \mu & \nu &  \lambda & \mu & \nu \\
        \mu & \nu & \lambda &  \lambda & \mu & \nu & \nu & \lambda & \mu \\
        \nu & \lambda & \mu &  \lambda & \mu & \nu &  \mu & \nu & \lambda \\
        \CodeAfter
        \OverBrace[shorten,yshift=1.5mm]{1-1}{1-3}{$r=0$}
        \OverBrace[shorten,yshift=1.5mm]{1-4}{0-6}{$r=1$}
        \OverBrace[shorten,yshift=1.5mm]{1-7}{0-9}{$r=2$}
               \SubMatrix{.}{1-9}{2-9}{\}}[xshift=3mm] 
               \SubMatrix{.}{3-9}{5-9}{\}}[xshift=3mm]
               \SubMatrix{.}{6-9}{8-9}{\}}[xshift=3mm] 
        \CodeAfter
        \tikz[remember picture,overlay]{
        \node[right=17pt] at ($(1-9.north east)!0.5!(2-9.south east)$) {$p=0$};
        \node[right=19pt] at ($(3-9.north east)!0.5!(5-9.south east)$) {$p=1$};
        \node[right=19pt] at ($(6-9.north east)!0.5!(8-9.south east)$) {$p=2$}; }
\end{pNiceMatrix}
\end{equation}

\noindent where $\lambda:=-i/\sqrt{3}$, $\mu:=\omega \lambda$ and $\nu:=\omega^2 \lambda$ when $\omega=e^{2 \pi i/d}$. The first column (blue) is the solution of saturation condition of SOS decomposition that depends on the choice of SICs. The other columns are generated by multipying the weight factor in the coefficient, $\omega^{-(ps+qr)}$, to the first column. The cyclic symmery underlying the weight is expressed in the block-structure of the coefficient matrix. Each block (white, gray) corresponds to a combination of $(p,r)$ and each element is determined further specifying $(q,s)$. The block matrices in the first row are obtained as a $(d-1)\times d = 2 \times 3$ matrix, as a trivial case $(p,q)=(0,0)$, or equivalently $dp+q\neq 0$ case, is excluded by assumption. One can find that all the blocks are one of the different distributions of three sets of $\lambda$, $\mu$, and $\nu$. 

We also present, for comparison, the coefficient matrix of Gisin's elegant Bell inequality (EBI), an example for the case $d=2$. In a similar manner, all the coefficients are determined by the multiplication of first column $(1/\sqrt{3})(-1,-1,1)$ derived in the main text and the weight factor, in this case $(-1)^{(ps+qr)}$. The matrix is therefore given by,\\

\begin{equation}
F_1^{EBI} = \frac{1}{\sqrt{3}}
\begin{pNiceMatrix}[margin] \Block[fill=blue!10]{1-1}{}
    -1 & -1 & \Block[fill=black!10]{1-2}{} 1 &1 \\ \Block[fill=blue!25]{2-1}{}
    -1 & 1 \Block[fill=black!10]{2-1}{} & -1 &1 \\ 
    1 & -1 & -1 &1 \\
        \CodeAfter
        \OverBrace[shorten,yshift=1.5mm]{1-1}{1-2}{$r=0$}
        \OverBrace[shorten,yshift=1.5mm]{1-3}{0-4}{$r=1$}
               \SubMatrix{.}{1-4}{1-4}{\}}[xshift=3mm] 
               \SubMatrix{.}{2-4}{3-4}{\}}[xshift=3mm]
        \CodeAfter
        \tikz[remember picture,overlay]{
        \node[right=17pt] at ($(1-4.north east)!0.5!(1-4.south east)$) {$p=0$};
        \node[right=17pt] at ($(2-4.north east)!0.5!(3-4.south east)$) {$p=1$}; }
\end{pNiceMatrix}
\end{equation}

\noindent The first column and the block-structure are denoted similarly to the above case. We note that the coefficient matrix is equivalent to that of Gisin's EBI up to a relabeling of Bob's measurement indices, which does not alter the equivalence between the Bell inequalities. Specifically, the reversion of the column order derives the case of EBI.

We next present the explicit forms of MUBs and SICs contained in the operator eigenspace corresponding to $S_3$. The maximal quantum violation, $\tilde{Q}=18$, of the Bell inequality $S_3$, is obtained from $\ket{\varphi_3^+}=(1/\sqrt{3}) 
(\ket{00}+\ket{11}+\ket{22})$, $A_x=W_x$, and $B_y=W_yB_0W_y^\dagger$ with $B_0:=\ket{0}\bra{0}+\omega\ket{\varphi_1}\bra{\varphi_1}+\omega^2\ket{\varphi_2}\bra{\varphi_2}$ when $\ket{\varphi_1}:=(1/\sqrt{2})(0,1,1)^\intercal$ and $\ket{\varphi_2}:=(1/\sqrt{2})(0,1,-1)^\intercal$. As the two eigenvectors of $B_0$, $\ket{\varphi_1}$ and $\ket{\varphi_2}$ are fiducial states, the sets of eigenvectors $\{W_y \ket{\varphi_1} \}$ and $\{W_y \ket{\varphi_2} \}$ respectively defines dual SICSs,
\begin{align*}
\mbox{SIC}_1&=
\frac{1}{\sqrt{2}}\begin{pmatrix}
    0&1&1&0        &\omega^2 &\omega   &0        &\omega   &\omega^2\\
    1&0&1&\omega   &0        &\omega^2 &\omega^2 &0        &\omega\\
    1&1&0&\omega^2 &\omega   &0        &\omega   &\omega^2 &0
\end{pmatrix} \,
\\
\mbox{SIC}_2&=
\frac{1}{\sqrt{2}}\begin{pmatrix}
    0&-1&1&0        &-\omega^2 &\omega   &0        &-\omega   &\omega^2\\
    1&0&-1&\omega   &0        &-\omega^2 &\omega^2 &0        &-\omega\\
    -1&1&0&-\omega^2 &\omega   &0        &-\omega   &\omega^2 &0 
\end{pmatrix}
\end{align*}
where the columns of $\mbox{SIC}_\beta$ correspond to the normalized vectors defining SICs generated with $\ket{\varphi_\beta}$. On the other side, the eigenstates of Alice's optimal observables $A_1=Z$, $A_3=X$, $A_4=XZ$, $A_5=XZ^2$ respectively define the complete mutually unbiased bases,
\begin{align*}
&\mbox{MUB}_1=
    \begin{pmatrix}
        1&0&0\\
        0&1&0\\
        0&0&1
    \end{pmatrix},\,
    \mbox{MUB}_2=
    \frac{1}{\sqrt{3}}\begin{pmatrix}
        1&\omega^2 &\omega\\
        1&\omega &\omega^2\\
        1&1&1
    \end{pmatrix}\\
&\mbox{MUB}_3=
    \frac{1}{\sqrt{3}}\begin{pmatrix}
        \omega^2 &\omega&1\\
        \omega^2 &1&\omega\\
        1&1&1
    \end{pmatrix},\,
    \mbox{MUB}_4=
    \frac{1}{\sqrt{3}}\begin{pmatrix}
        \omega&1&\omega^2\\
        \omega&\omega^2&1\\
        1&1&1
    \end{pmatrix}.
\end{align*}
As discussed in Section I, the eigenvectors of other four operators of Alice, also form the above complete MUBs.


\begin{thebibliography}{99}
\bibitem{Note}
The index takes the nontrivial value $\tau=(i)^{pq}$ only for the case, $d=2$, otherwise $\tau=1$. It is introduced to satisfy the correct spectrum condition of Alice's observable when $d=2$, i.e. $A_3^2=\mathds{1}$. This special treatment is needed for $d=2$ as it is the only even prime dimension.
\end{thebibliography}
\end{document}